\documentclass[%
 reprint, 
groupedaddress,
 preprintnumbers,
 showkeys,
 amsmath,amssymb,
 aps,
]{revtex4-2}

\usepackage{xcolor}
\definecolor{darkblue}{RGB}{46,48,147}
\usepackage{url}

\usepackage{breakurl}
\usepackage[colorlinks=true,
            linkcolor=darkblue,
            urlcolor=darkblue,
            citecolor=darkblue,
            breaklinks=true]{hyperref}
\usepackage{xspace}
\usepackage{graphicx}
\usepackage{dcolumn}
\usepackage{bm}
\usepackage[normalem]{ulem}  

\graphicspath{{figures/}}

\definecolor{darkbluegrey}{rgb}{0.45, 0.43, 0.66}
\definecolor{vermillion}{rgb}{0.86, 0.18, 0.01}

\newcommand{\ms}[1]{\textsf{\color{blue}{\textsuperscript{MS}#1}}}

\newcommand{\BVIMIN}{\texttt{B:VIMIN}\xspace}

\newcommand{\BIMINER}{\texttt{B:IMINER}\xspace}
\newcommand{\BLINFRQ}{\texttt{B:LINFRQ}\xspace}

\newcommand{\IIB}{\texttt{I:IB}\xspace}
\newcommand{\IMDATFORTY}{\texttt{I:MDAT40}\xspace}

\usepackage{scalerel}
\usepackage{tikz}
\usetikzlibrary{svg.path}

\definecolor{orcidlogocol}{HTML}{A6CE39}
\tikzset{
  orcidlogo/.pic={
    \fill[orcidlogocol] svg{M256,128c0,70.7-57.3,128-128,128C57.3,256,0,198.7,0,128C0,57.3,57.3,0,128,0C198.7,0,256,57.3,256,128z};
    \fill[white] svg{M86.3,186.2H70.9V79.1h15.4v48.4V186.2z}
                 svg{M108.9,79.1h41.6c39.6,0,57,28.3,57,53.6c0,27.5-21.5,53.6-56.8,53.6h-41.8V79.1z M124.3,172.4h24.5c34.9,0,42.9-26.5,42.9-39.7c0-21.5-13.7-39.7-43.7-39.7h-23.7V172.4z}
                 svg{M88.7,56.8c0,5.5-4.5,10.1-10.1,10.1c-5.6,0-10.1-4.6-10.1-10.1c0-5.6,4.5-10.1,10.1-10.1C84.2,46.7,88.7,51.3,88.7,56.8z};
  }
}

\newcommand\orcidicon[1]{\href{https://orcid.org/#1}{\mbox{\scalerel*{
\begin{tikzpicture}[yscale=-1,transform shape]
\pic{orcidlogo};
\end{tikzpicture}
}{0}}}}
\begin{document}

\title{Uncertainty Aware ML-based surrogate models for particle accelerators: A Study at the Fermilab Booster Accelerator Complex}

\author{Malachi~Schram\,\orcidicon{0000-0002-3475-2871}}
\email{schram@jlab.org} 
\author{Kishansingh Rajput\,\orcidicon{0000-0002-4430-9937}}
\affiliation{Thomas Jefferson National Accelerator Laboratory, Newport News, VA 23606, USA}
\author{Karthik Somayaji NS\,\orcidicon{ 0000-0002-6937-8082 }}
\author{Peng Li\,\orcidicon{ 0000-0003-3548-4589 }}
\affiliation{University of California Department of Electrical and Computer Engineering, Santa Barbara, CA, 93106, USA}

\author{Jason~St.~John\,\orcidicon{0000-0001-8110-4108}}
\affiliation{Fermi National Accelerator Laboratory, Batavia, Illinois 60510, USA}

\author{Himanshu Sharma\,\orcidicon{0000-0001-5396-1276}}
\affiliation{Pacific Northwest National Laboratory, Richland, WA 99354, USA}

\begin{abstract}
Standard deep learning methods, such as Ensemble Models, Bayesian Neural Networks and Quantile Regression Models provide estimates to prediction uncertainties for data-driven deep learning models.
However, they can be limited in their applications due to their heavy memory, inference cost, and ability to properly capture out-of-distribution uncertainties.
Additionally, some of these models require post-training calibration which limits their ability to be used for continuous learning applications.
 In this paper, we present a new approach to provide prediction with calibrated uncertainties that includes out-of-distribution contributions and compare it to standard methods on the Fermi National Accelerator Laboratory (FNAL) Booster accelerator complex.
\end{abstract}

\maketitle

\section{Introduction}\label{sec:intro}
Particle accelerators are complex multi-system machines that include a large number of variables with non-linear dynamics.
To date, accelerator control systems are manually optimized by experts that are guided by physical principles whenever possible.
Developing high-dimensional, physics-based models that account for multiple time scales is extremely challenging. Although, there are accelerator beam models with impressive and improving precision~\cite{synergia}, building fully comprehensive Monte Carlo-based models of an entire facility is challenging, if not intractable.
Data-driven methods, such as deep neural networks (DNNs), are well suited to capture the dynamics of these non-linear complex systems. 
These surrogate models can then be used to develop new AI-based control systems provided they can inform the optimization algorithm on how reliable the predictions are.
\\ \\
The recent development of DNNs~\cite{LeCun2015,Goodfellow-et-al-2016,Carleo:2019ptp} has proven itself useful for complex control problems~\cite{deepcool,wiredcars,microsoftai,mitfanuc}.
The use of machine learning for particle accelerator applications has grown in recent years to include, but is not limited to, diagnostics~\cite{Emma:2018meg, Sanchez-Gonzalez:2016zhm,Wielgosz:2016xhl,Scheinker:2015mra, alex2020advanced, PhysRevLett.121.044801,Li_2021}, anomaly detection/forecasting/classification~\cite{Blokland:2021onk,Rescic:2020ueu,RESCIC2022166064,tennant2020superconducting,Powers:2019ioo}, and controls~\cite{PhysRevAccelBeams.24.104601,Kafkes:2021jse, Edelen:2016dqu, hirlaender2020modelfree}.
Although these studies have shown some impressive results, the majority do not include any uncertainty quantification (UQ) to complement their predictions.
Unfortunately, the use of DNNs for online safety-critical applications remains limited due to issues such as model explainability, in-domain and out-of-domain prediction and uncertainties, and uncertainty calibrations.\\ \\
In recent years, there has been an increasing amount of effort on estimating uncertainties in DNNs. 
A prediction's uncertainty can be separated into the model's intrinsic uncertainty (model uncertainty) and the uncertainty caused by the data (data uncertainty). The model uncertainty is typically reducible, within limits, by improving the model architecture and hyper-parameters, however, the data uncertainty is irreducible.
Additionally, uncertainty estimation originating from Out-Of-Distribution (ODD) samples is critical for a number of applications, such as using DNNs as a proxy to model a dynamical system used for system control and/or optimization. 
A deep learning method that provides predictive uncertainty is not sufficient for safe decision-making; a deep learning method with \textit{properly calibrated} uncertainty is required.\\ \\
Recent studies that include data-driven, Machine Learning (ML) based surrogate models have started to include UQ in their models, such as developing a UQ-based surrogate model of cyclotron-based model~\cite{Andreas2019},  modeling the FNAL Booster accelerator complex for a reinforcement learning application~\cite{Kafkes:2021jse}, and on uncertainty aware anomalies prediction~\cite{Blokland:2021onk}.
Additionally, a recently published study compared the use of Bayesian Neural Networks (BNN) and ensemble methods for particle accelerator applications~\cite{PhysRevAccelBeams.24.114601}. 
In this paper, we compare three different methods to estimate data-related uncertainties for DNN models as it applies to modeling the FNAL Booster Accelerator Complex.
In Section~\ref{sec:complex}, we briefly describe the Fermilab Booster accelerator complex and the data used for training the DNN models in the  context of a control optimization problem. 
In Section~\ref{sec:methods}, we introduce three methods that estimate uncertainty quantification for DNNs.
In Section~\ref{sec:results}, we present the performance of each method for in-distribution and out-of-distribution scenarios.
Finally, we conclude with a summary of our results in Section~\ref{sec:summary}.

\section{Fermilab Booster Accelerator and Complex}
\label{sec:complex}
\begin{figure}
\centering
\includegraphics[width=0.42\textwidth]{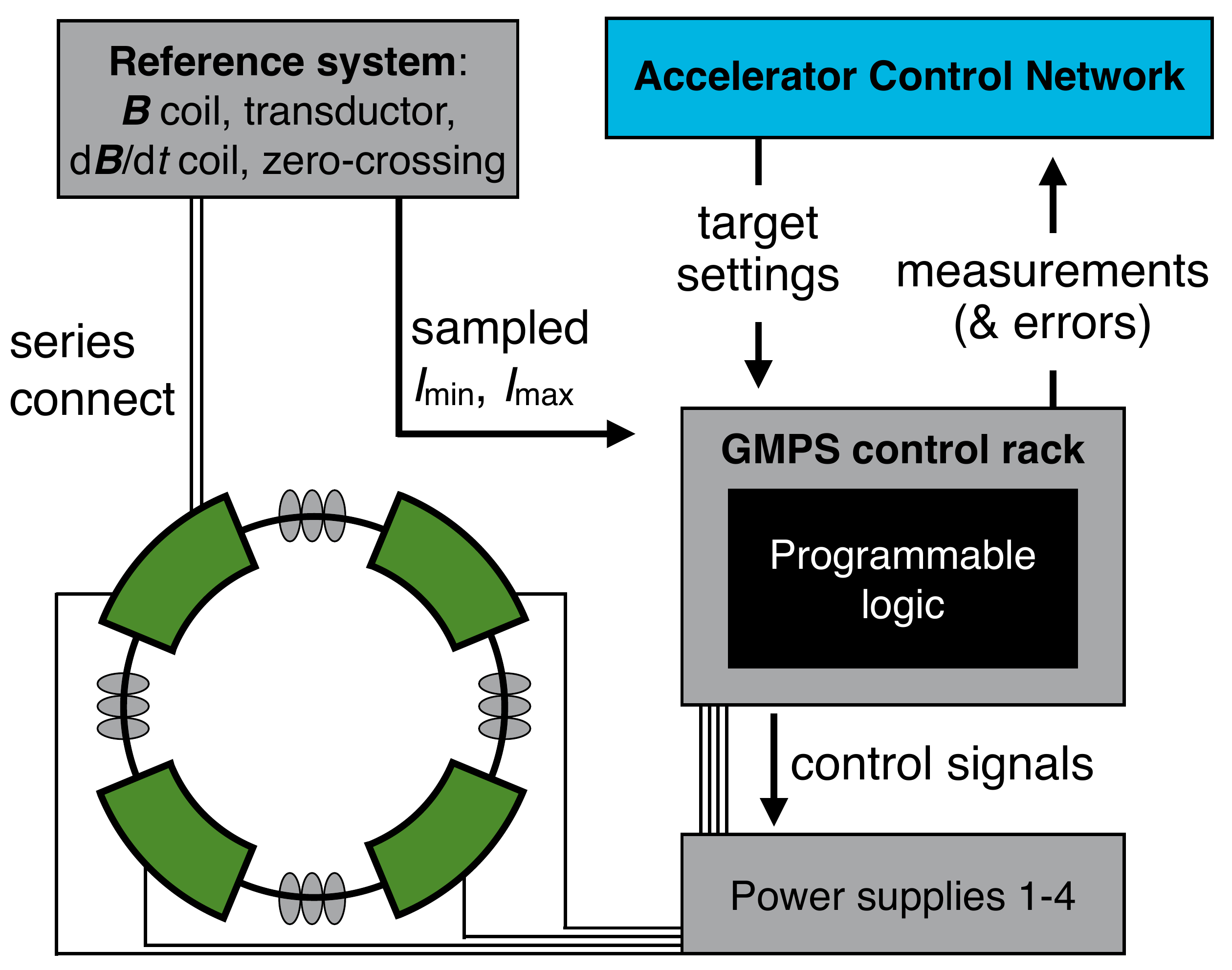}
\caption{Schematic view of the GMPS control environment.
The human operator specifies a target program via the Accelerator Control Network that is transmitted to the GMPS control board.
The FPGA-based control logic utilizes these settings together with readings from a reference magnet to prescribe a driving signal to the GMPS.
The effect of this prescribed signal on the bending magnets is measured by an in-series reference magnet, with sampled readings transmitted back to the GMPS control board.
Reference measurements and prescribed signals may be logged and transmitted over network for later analysis.}
\label{fig:schematic}
\end{figure}

\begin{figure}
\centering
\includegraphics[width=0.49\textwidth]{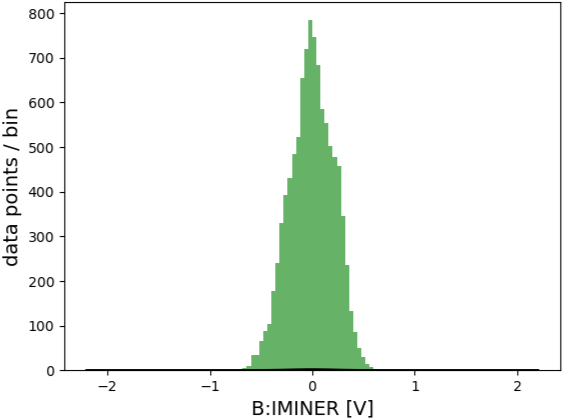}
\caption{Distribution of fractional measured error in the GMPS current at the minimum value of the magnet current (prefix \textbf{B:} indicates Booster, \textbf{I} current, \textbf{MIN} miniumum, and \textbf{ER} error), with the non-ML PID regulator discussed in the text. From~\cite{PhysRevAccelBeams.24.104601}.}
\label{fig:errdist}
\end{figure}

At 15~Hz the Fermilab Booster rapid-cycling synchrotron accelerates each injected batch of 400~MeV protons to 8~GeV and resets to receive the next injection. 
See \cite{PhysRevAccelBeams.24.104601} for detailed discussion.
A central component of the Booster cycle is the Gradient Magnet Power Supply (GMPS), which provides the synchronously rising and falling electrical current to this circular synchrotron's main bending magnets, tracking the energy (and therefore the proton beam's magnetic stiffness) upward to extraction before returning to the injection state. \\ \\
The throughput efficiency of the synchrotron is sensitive to unwanted perturbations of the GMPS current, causing the beam's trajectory to deviate from the desired path and scrape on apertures.
Such perturbations are understood to be induced by the power supplies of other nearby synchrotrons on their own cycles in the accelerator complex, temperature variations, 60~Hz power line frequency meanders, and other accelerator complex nuances.
A proportional\textendash integral\textendash derivative (PID~\cite{pid1,pid2}) regulator circuit attempts to compensate for these perturbations with cycle-by-cyle adjustments to the minimum of the sinusoidal control signal.
Figure~\ref{fig:schematic} shows a schematic overview of the GMPS control environment.  See Figure~\ref{fig:errdist} for a sample distribution of measured errors for the minimum value of the sinusoidally varying magnet current.\\ \\
Machine Learning techniques promise a new avenue for developing more sophisticated control agents with better overall regulation performance, allowing a predictive, anticipatory approach not encompassed by the reactive PID regulation paradigm.
As the authors in \cite{PhysRevAccelBeams.24.104601} point out, any ML-based GMPS regulator which replaces this PID circuit is required to deliver stable, fast inference times; the data intake, forward inference, and generation of the resulting control signal must always be complete in less than 66 milliseconds.\\ \\
Reinforcement Learning (RL) was selected as the approach to train a ML-based agent to act as the GMPS regulator. RL learns an action policy by training a model using data describing a system's states, actions, and the resulting outcomes. 
This technique is a natural choice because, once a competent agent is in operation, its real-world performance can be used to provide updated model parameters, tracking the slowly changing dynamics of the accelerator complex.
At the outset of RL training, the control agent would be expected to make egregious mistakes and to learn from them.
Thus a surrogate model is needed, one which captures the dynamics of the Booster GMPS regulator's control environment, where the agent can learn from mistakes without risk to personnel, equipment, or the science program they support around the clock. \\ \\
Surrogate model training and testing data were taken with the PID regulation circuit in operation. 
Additionally a small amount of data were taken with the regulation circuit off, or with changes to its coefficients. 
This choice maximized the available volume of training and testing data, sampling the changing response dynamics of the GMPS regulation while minimizing impact on accelerator operations. 
The PID regulator circuit's residual error is typically only 0.1\% of the injection current minimum. 
Without regulation, the fitted minimum of the magnetic field may vary from the set point by as much as a few percent. \\ \\
Five time series were used to produce the surrogate model, and we use their names as they are logged by the accelerator control system. 
(In the accelerator control system's data-logging nomenclature, device parameters with the \texttt{B:} prefix are related to the Booster, whereas device parameters beginning with \texttt{I:} are related to the Main Injector. 
``MDAT" denotes the accelerator (``machine") data communication broadcast.)
\BVIMIN is the compensating recommendation for the minimum value of the offset-sinusoidal GMPS current, issued by the GMPS regulator in order to reduce the magnitude of \BIMINER, which is a measure of the residual error. 
\BLINFRQ is the measured offset from the expected 60~Hz line frequency powering the GMPS, in mHz. 
\IIB and \IMDATFORTY provide measurements of the Main Injector bending dipole current at different points in the circuit and through different communication channels. 
Among all candidate device time series analyzed, Main Injector's power supplies were shown in \cite{PhysRevAccelBeams.24.104601} to have the highest Granger~\cite{granger} causality with respect to \BIMINER, the minimization control objective of this application, whose typical value distribution is shown in Figure~\ref{fig:errdist}.

Table~\ref{tab:parameters} briefly summarizes the parameters of interest used in surrogate modeling.\\ \\

\begin{table}[ht!]
\caption{Description of dataset parameters chosen by experts and later validated with a causality study. 
Here,``MI" means Main Injector, ``MDAT" means accelerator (machine) data communication, and device parameters that begin with \texttt{B} are related to the Booster, whereas device parameters that begin with \texttt{I} are related to the Main Injector. (Reused with permission from the authors of~\cite{PhysRevAccelBeams.24.104601})}

\centering
\resizebox{\columnwidth}{!}{
\begin{ruledtabular}
\begin{tabular}{l|l}
Parameter & Details [Units] \\ \hline
\BIMINER & Setting-error discrepancy at injection [A] \\ 
\texttt{B:LINFRQ} & 60~Hz line frequency deviation [mHz] \\ 
\BVIMIN & Compensated minimum GMPS current [A] \\ 
\IIB & MI lower bend current [A] \\ 
\IMDATFORTY & MDAT measured MI current [A] \\ 
\end{tabular}
\end{ruledtabular}}
\label{tab:parameters}
\end{table}

Data were collected nominally at 15~Hz, and due to clock drift among the front-ends taking those samples, the data as logged were then time-aligned to a periodic reference signal.
Period 0 (June 3, 2019 to July 11, 2019) was ended by the annual Summer Shutdown and Maintenance. Period 1 (December 3, 2019 to April 13, 2020) ended when the accelerator operations were suspended in response to the COVID-19 pandemic.
More detail on the collection and preparation of the data can be found at the Data Descriptor article~\cite{BOOSTR:datasheet}.

\section{Machine Learning Methods}
\label{sec:methods}
There has been a lot of research in uncertainty quantification for deep learning models that includes, but not limited to, BNN~\cite{gal2016dropout}, Deep Quantile Regression (DQR)~\cite{koenker_2005}, and Deep Gaussian Process Approximation models (DGPA) \cite{NIPS2007_013a006f,Rasmussen2004,NEURIPS2020_543e8374,DKL}. 
In this paper, we do not consider ensemble methods because these methods require training of multiple models and multiple inferences to provide an uncertainty estimation, making it computationally expensive, slow, and memory intensive.
For this paper, we implemented models for BNN, DQR, and DGPA to better understand their performance for in-distribution and out-of-distribution uncertainty estimation and report the results in the next section.
Our specific effort to develop a new DGPA model is most closely related to the recent paper on Spectral-normalized Neural Gaussian Process (SNGP) models~\cite{SNGP2020} for classification. 
In the following subsections we discuss the working mechanism of these methods.
\subsection{Bayesian Neural Network (BNN) Model}
Monte-Carlo (MC) Dropout~\cite{gal2016dropout} is a commonly used approach to estimate the prediction uncertainty for deep neural network models. 
The MC-dropout approach has shown~\cite{gal2016dropout} to overcome the computational challenges of estimating uncertainty using Bayesian models. 
The review by Abdar et.al~\cite{abdar2021review} provide a  comprehensive details on estimating uncertainty in deep learning. The initial application of MC dropout was to over-come over fitting associated with deep neural networks (DNN) while training. \\ \\
The MC dropout approach relies on introducing a tunable uncertainty into a network training process by adding a dropout layer that randomly removes nodes to the following layer with a set probability ($p$) at each forward pass in training process. 
While training the DNN with dropout, the units in a layer are randomly dropped is typically used to avoid over-fitting, it can also be used during inference to estimate the uncertainties.
In order to combing the aleatoric and epistemic uncertainty into the BNN model, we implemented the loss function described in~\cite{kendall2017}, as shown in Equation~\ref{eq:LossDrop}.
\begin{equation}\label{eq:LossDrop}
\mathcal{L}_{\text {dropout }}=\frac{1}{N} \sum_{i=1}^{N} \frac{1}{2\sigma(x_{i})^{2}}||\mathbf{y}_{i} - \mathbf{y}^{p}_{i}||^{2} + \frac{1}{2}\log\sigma(x_{i}^{2})
\end{equation}
With this loss function, we can account for both sources of uncertainties and provide a calibrated model.
Here $y^{p}_i$ and $y_i$ are the the prediction and measured values, and $\sigma$ is the model’s predicted noise which is determined during training by optimizing the dropout level to minimize Equation~\ref{eq:LossDrop}.
\subsection{Deep Quantile Regression (DQR) Model }
DQR is a method used to estimate the conditional quantiles of a response variable distribution that is more robust against outliers in the response measurements~\cite{koenker_2005}. 
We define the conditional quantile function for a non-linear relationship in Equation~\ref{eq:quantile}.
\begin{equation}
Q_{y}(\tau|x_{t}) = G_{\tau}(x_{t}, w)
\label{eq:quantile}
\end{equation}
Here $G_\tau(x_{t}, w)$ is a non-linear function that is approximated by a DNN, $x$ is the input feature vector at time $t$, and $\tau^{th}$ is the conditional quantile.
We develop the DNN model to simultaneously learn predictions based on a set of defined quantiles.
For each defined quantile, the prediction $y_{i}^{p}$ and outcome $y_{i}$, the regression loss for a $\tau^{th}$ quantile is given in Equation~\ref{eq:QRloss}:
\begin{equation}
\mathcal{L}(\mathbf{y_{i}^{p}},\mathbf{y_{i}}) = \max[\tau(\mathbf{y}_{i}-\mathbf{y}_{i}^{p
}),(\tau-1)(\mathbf{y}_{i}-\mathbf{y}_{i}^{p})]
\label{eq:QRloss}
\end{equation}
As such, DQR provides a comprehensive statistical model that captures non-linear relationships by providing conditional quantiles along with the median, in contrast to traditional regression methods.
In this paper, we implemented a deep learning model with convolutional layers, similar to the model architecture used for the other methods discussed in this paper, however, the output layers mapping to a dedicated quantile value. 
\subsection{Deep Gaussian Process Approximate (DGPA) Model}
Gaussian Process (GP) models uses a kernel function to transform the input data into some higher dimensional representation. 
The function utilizes the point-to-point distance between samples in the new representation to produce predictions. 
With this property it's intrinsically distance aware and can detect OOD samples based on the distance from training distribution. 
Unfortunately, GP models do not scale with large data sets and large feature dimensions.
As such, using GP on high dimensional data usually requires either dimension reduction, feature extraction or some other form of approximation.
In contrast, DNNs can be very expressive and readily applied to problems with large data sets and high dimensional feature space. 
Unfortunately, deterministic DNNs can make predictions on samples that are outside their training data set that are not guaranteed to be accurate~\cite{NIPS2009_e7f8a7fb} and are unable to identify these predictions as being OOD. 
As such, we incorporated the desired qualities of the GP into the DNN by adding a fixed size lower rank approximation of the GP with an RBF kernel, $K = \Phi \Phi^{T}$, at the final layer using random fourier features (RFF) as defined in \cite{https://doi.org/10.48550/arxiv.2006.10108}.
Although this provides a uncertainty estimation that is distance aware, it is not distance preserving because there is no guarantee that distance between the input data is preserved at the hidden layer where the GP approximation is applied. 
In order to make the DNN distance preserving, we used the \textit{bi-Lipschitz} constraint as part of the training loss function, as shown in Equation~\ref{eq:biLips}.
\begin{equation}\label{eq:biLips}
L_{1} \times ||x_{1} - x_{2}|| \leq ||h_{x_{1}} - h_{x_{2}}|| \leq L_{2} \times ||x_{1} - x_{2}||
\end{equation}
Here $x$ is the input feature vector and $h_{x}$ be the last hidden layer output.
To summarize, we implemented the same model architecture used for the other methods in this paper, however, we introduced a GP RBF kernel approximation with 256 RFF and modified the loss function to ensure the distance between input and hidden layers are preserved using soft \textit{bi-Lipschitz} constraint using $L_{1}=0.75$ and $L_{2}=1.25$. 
\section{RESULTS}
\label{sec:results}
\begin{figure}
\centering
\includegraphics[width=0.49\textwidth]{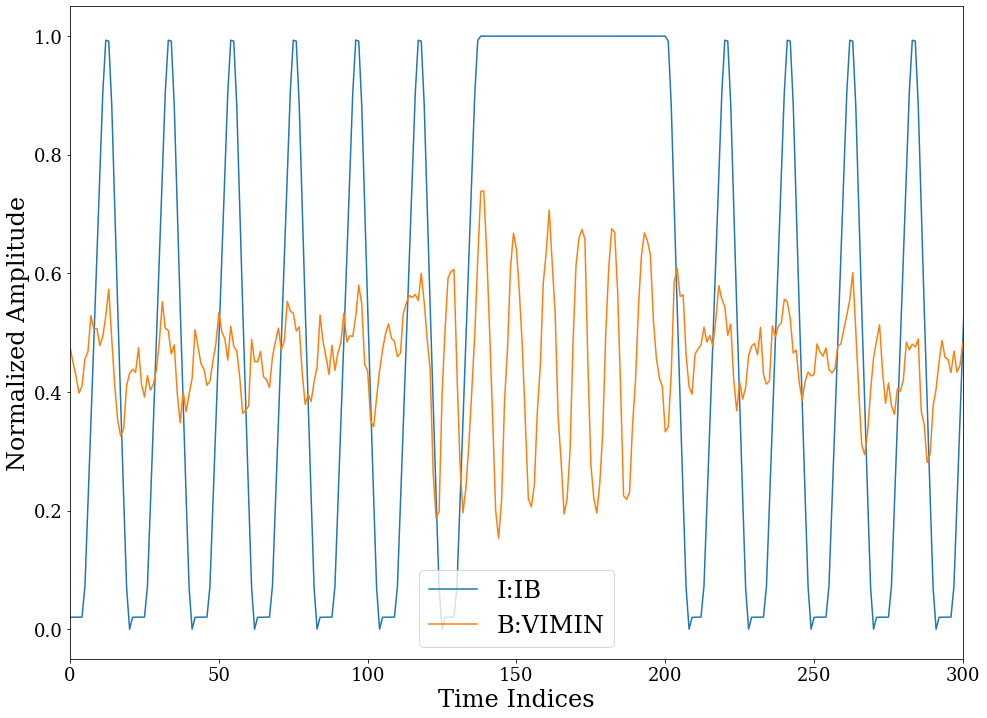}
\caption{Comparison between the main injector lower bend current and the compensated minimum GMPS current.}
\label{fig:InjectorCut}
\end{figure}

Traditional deep learning models are deterministic and provide a prediction for each input with no measure of confidence associated with the prediction. 
Providing methods with reliable predictive uncertainty for ML models is critical for real-world applications. 
As discussed in the previous section, there are number of methods being proposed in the literature to make the DL models uncertainty aware. 
In this section we compare the performance of the methods presented in Section~\ref{sec:methods} for in-distribution and out-of-distribution samples as it applies to the prediction for the FNAL Booster accelerator complex.
The input samples that are independent and identically distributed (\textit{iid}) as training data (in-distribution), if the underlying system/data produces noisy labels, a DL model will learn to produce the mean. 
We expect the uncertainty values for such predictions to reliably represent the variance in the underlying data labels. 
The input samples that are dissimilar to the training samples, called OOD samples, are most difficult for a DL model to provide the prediction accurately; most of the time, the model will produce inaccurate predictions leading to unreliable results. A prediction without the associated calibrated, distance-aware uncertainty quantification results in a system without contextual information required to select a safe response for a prediction.
We require the uncertainty values for OOD predictions to be high indicating low confidence in the prediction.\\ \\
To compare the results, we trained all the models with similar architecture and data sets. The models consist of three convolutional blocks where each block contained a 1-dimensional convolutional layer with 32 filters of size 3 followed by a batch-normalization, a maxpooling layer, and a dropout layer with probability of 0.1.  The output of the third convolutional block is then flattened to process through a dense layer containing 256 nodes leading to the output layer.
The differences between the three models are in how they quantify uncertainty.
The DGPA model has a Gaussian approximation layer as the output layer, BNN has a vanilla dense layer as output layer but the dropouts are kept on during inference, and DQR has multiple dense layers to produce output for different quantiles. \\ \\
For this study, the raw data were processed using MinMax scaling and restructured so that 15 previous timesteps of the input variables were fed into the models to predict the next timestep forward in the output variables.
We divided the data into orthogonal samples, $80\%$ for training and $20\%$ for testing.
The samples were further filtered by explicitly excluding contributions when the main injector lower bound current (\IIB)  had a value that exceeded 0.995.
This filter was used to create a in distribution only training samples that would prevent the models to see the cyclic high amplitude in the predicted variable (\BVIMIN).     
The relationship between the filtered variable (\IIB) and the predicted variable (\BVIMIN) is shown in Figure~\ref{fig:InjectorCut}.
\subsection{In-distribution results}\label{sec: in-distribution-results}
\begin{figure}
\centering
\includegraphics[width=0.4\textwidth]{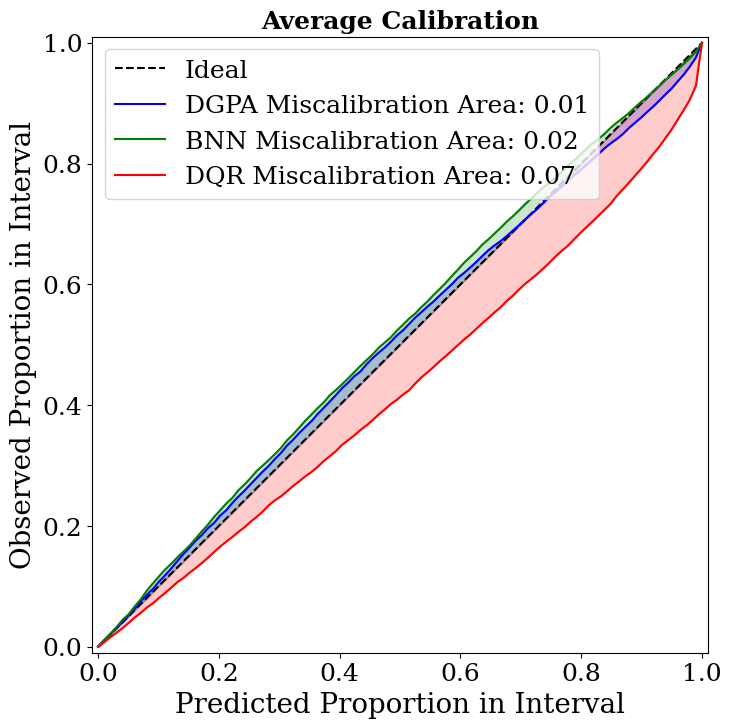}
\caption{Comparison of miscalibration between DGPA, BNN and DQR for in-distribution data. The shaded area represents the amount of miscalibration in the uncertainty estimation with respect to the true labels.}
\label{fig:id_miscalibrationPlot}
\end{figure}

For in-distribution input samples, deep learning models are expected to have accurate predictions with an uncertainty estimation  that is consistent with the training data.
To compare the predictive performance along with the uncertainty quantification for these models we use a set of standard metrics including R-square, Root Mean Square Error (RMSE) between the ground truth labels and the predictions, Mean Absolute Calibration Error (MACE), and Root Mean Square Calibration Error (RMSCE) from the uncertainty toolkit~\cite{chung2021uncertainty}. 
Table~\ref{tab:fit-results} shows the values of these metrics for each of the three models before performing any calibration. \\ \\
All three models have very similar predictive performance in terms of $R^2$ and RMSE and their uncertainty estimations as shown in Figure \ref{fig:id_miscalibrationPlot} and Table~\ref{tab:fit-results}. 
\begin{table}[ht!]
\caption{In distribution prediction performance for BNN, DQR, and DGPA models. \\ \\
}
\centering
\resizebox{\columnwidth}{!}{
\begin{ruledtabular}
\begin{tabular}{l|l|l|l|l}
Model & $R^2$ & RMSE & MACE & RMSCE  \\ \hline
BNN & 0.835 & 0.032 & 0.020 & 0.022  \\ 
DQR & 0.864 & 0.029 & 0.015 & 0.017 \\ 
DGPA & 0.853 & 0.030 & 0.014 & 0.016 \\ 
\end{tabular}
\end{ruledtabular}}
\label{tab:fit-results}
\end{table}

\subsection{Out-of-distribution results}
\label{sec: ood-results}
The majority of ML applications assumes that the test samples for a model are \textit{iid} and similar to the training data. 
Unfortunately, in practice this assumption doesn't always hold. 
When a test data draws from an out of training distribution sample, the trained model is not guaranteed to produce accurate predictions\cite{NIPS2009_e7f8a7fb}. 
Providing an uncertainty estimation consistent with the distance between in-distribution and out-of-distribution data is desirable.
We expect the model to produce high uncertainty for inaccurate predictions and lower uncertainty when the predictions are accurate. We expect this relationship between uncertainty and error even for the OOD samples.
To evaluate the ability of each model's technique to estimate the OOD uncertainty, we created two scenarios detailed below.

\subsubsection{Scenario 1}\label{sec:scenario1}
\begin{figure}
\centering
\includegraphics[width=0.4\textwidth]{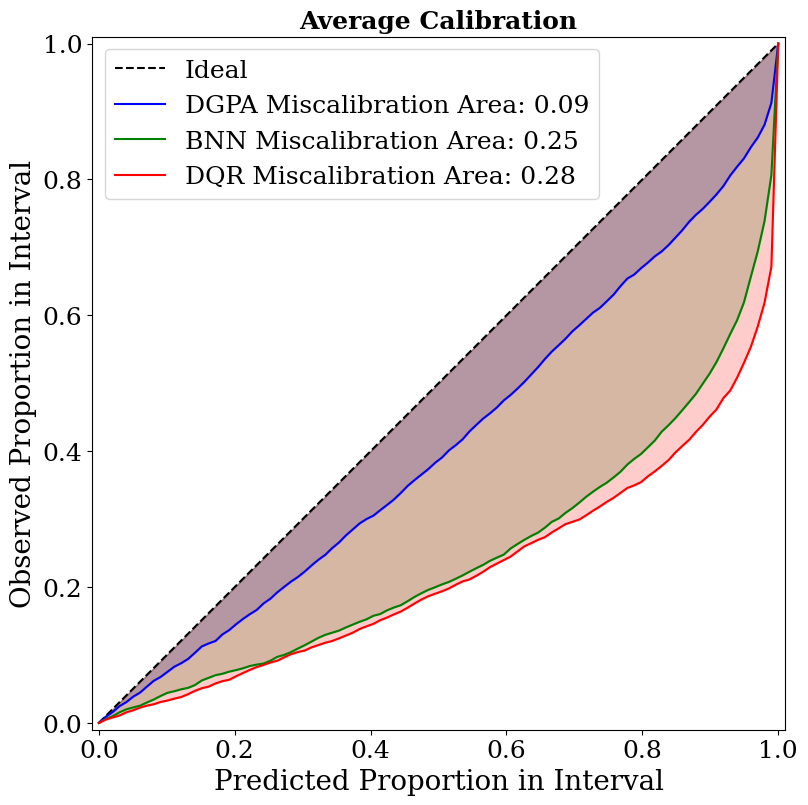}
\caption{Comparison of miscalibration between DGPA, BNN and DQR for OOD data. The shaded area represents the amount of miscalibration in the uncertainty estimation with respect to the true labels}
\label{fig:miscalibrationPlot}
\end{figure}

For the first scenario, we trained the models with the in-distribution samples and evaluated their performance using the full test sample. 
\begin{figure}
\centering
\includegraphics[width=0.49\textwidth]{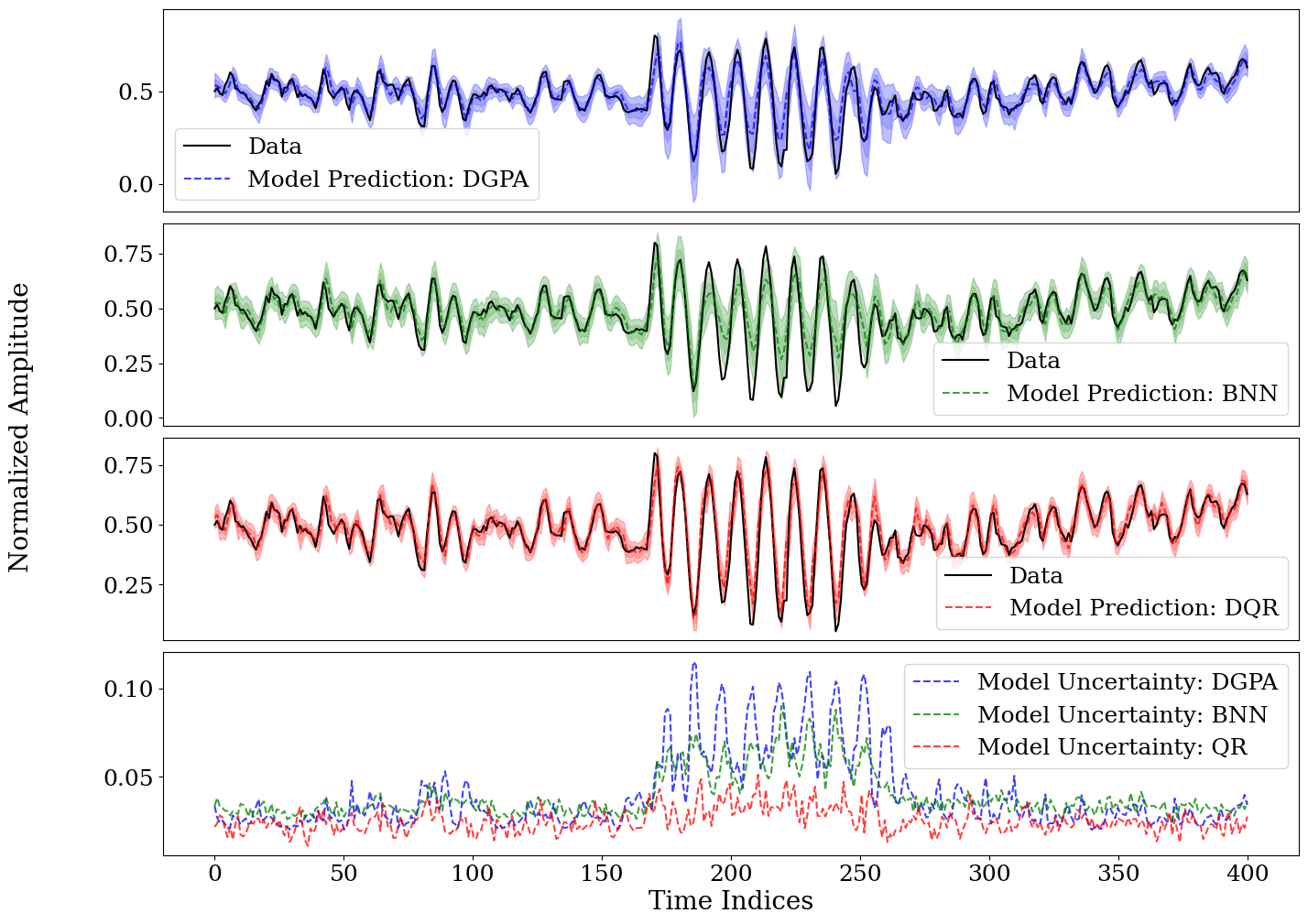}
\caption{Side by side comparison of the predictive performance as well as uncertainty quantification results for in-distribution and out-of-distribution samples for DGPA, BNN, and DQR respectively. The middle region with the high frequency component on the time series represent OOD samples while the initial and tail-end regions represent in-distribution data samples.}
\label{fig:OOD_real}
\end{figure}
The top three plots in Figure \ref{fig:OOD_real} show each model's predictions and uncertainties using data that includes the OOD samples.
As can be seen, the predictions from all three models degrades in the OOD region, which is expected. 
The respective uncertainty values are expected to correlate to the deviation from the ground truth labels.
We used the same metrics from the in-distribution study to quantify how close each model is to the ground truth and present the results in Table~\ref{tab:fit-odd_results}.
These results show how close the models are to the ground truth and is not used as a calibration study.
The $R^2$ and RMSE is similar for all three models, however, the uncertainty estimation varies. 
As shown in Figure~\ref{fig:miscalibrationPlot}, all models underestimate the uncertainty, however, both the uncertainty predictions from the BNN and DQR are significantly lower than those from the DGPA.\\ \\
The DPGA has approximately 3x smaller MACE and RMSCE than the DQR and BNN which indicated that the overall  uncertainty estimation of the predictions are closer to the ground truth. 
\begin{table}[ht!]
\label{tab:OOD}
\caption{Out of distribution prediction performance for BNN, DQR, and DGPA models. \\ \\
}
\centering
\resizebox{\columnwidth}{!}{
\begin{ruledtabular}
\begin{tabular}{l|l|l|l|l}
Model & $R^2$ & RMSE & MACE & RMSCE\\ \hline
BNN & 0.694 & 0.104 & 0.252 & 0.281 \\ 
DQR & 0.836 & 0.076 & 0.275 & 0.309 \\ 
DGPA & 0.784 & 0.088 & 0.091 & 0.100\
\ 
\end{tabular}

\end{ruledtabular}}
\label{tab:fit-odd_results}
\end{table}
\subsubsection{Scenario 2}\label{sec:scenario2}
\begin{figure}[t]
\centering
\includegraphics[width=0.48\textwidth]{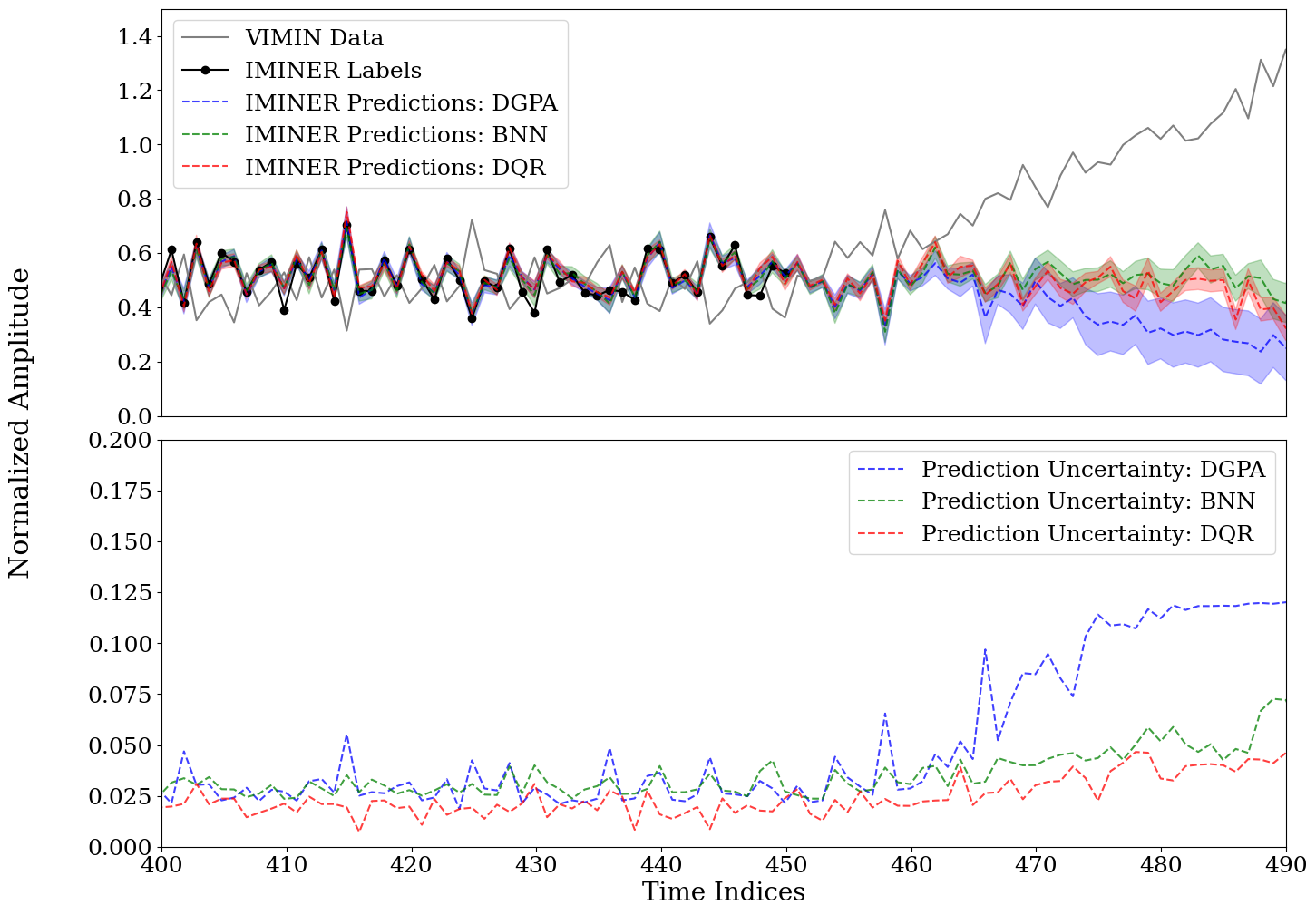}
\caption{Comparison of the predictive performance with uncertainty quantification between DGPA, BNN, and DQR respectively for a manually induced OOD. The vertical line at 450 indicates the area where the VIMIN input variable was incrementally increasing to induce the OOD scenario.}
\label{fig:OOD_2}
\end{figure}

Figure~\ref{fig:OOD_2} shows the second scenario where we monotonically increase one of the key input variable VIMIN until the data samples enters into a region of feature space that is out of training distribution. 
VIMIN is chosen because it is one of the key variable affecting our target.
These data samples are then fed into the models for inference and uncertainty quantification. Since these data samples are not within the training distribution they are seen as OOD by the models. 
Similar to the above discussed OOD results, the uncertainty values from the BNN and DQR are underestimated as compared to DGPA when the data samples goes into OOD region.
From Figure \ref{fig:OOD_real} and Figure \ref{fig:miscalibrationPlot} as well as the Figure \ref{fig:OOD_2} it is clear that for OOD samples, DGPA produces more accurate uncertainty estimates as compared to BNN and DQR, both of which underestimate the uncertainty values.
\section{Summary and Conclusions}\label{sec:summary}
In this paper, we compared three different DNN techniques that estimate the prediction uncertainty. 
We present the DGPA technique as a new approach to estimating prediction uncertainties; DGPA is self-calibrated and includes an awareness of out-of-distribution uncertainty. 
The results show that all models provide similar performance for the predictions for in-distribution, however, the DGPA provides uncertainty estimations that are closer to the ground truth for the out-of-distribution Scenario 1. 
Although we cannot quantify the explicit expected uncertainty for Scenario 2, we can see that the uncertainty estimation from the DPGA is larger than the other two methods where the DQR provides a very small uncertainty estimation.
In conclusion, the results from this study on the FNAL Booster Accelerator Complex data suggestion that the DGPA model provides the best single inference calibrated data driven ML-based model for in and out of distribution uncertainty estimation for all scenarios.  
This was achieved by using a fixed size GP RBF kernel approximation and applying the \textit{bi-Lipschitz} constraint in the loss function.
Additional research should be conducted to better understand the trade-off from kernel approximation size and how this can be applied to real-time systems where the hardware could be a constraint.
\begin{acknowledgments}
This research was supported by the U.S. Department of Energy, through the Office of Advanced Scientific Computing Research's “Data-Driven Decision Control for Complex Systems (DnC2S)” project. Pacific Northwest National Laboratory is operated by Battelle Memorial Institute for the U.S. Department of Energy under Contract No. DE-AC05-76RL01830. Oak Ridge National Laboratory is operated by UT-Battelle LLC for the U.S. Department of Energy under contract number DE-AC05-00OR22725. The Jefferson Science Associates (JSA) operates the Thomas Jefferson National Accelerator Facility for the U.S. Department of Energy under Contract No. DE-AC05-06OR23177.
\end{acknowledgments}

\bibliography{references}

\end{document}